\documentclass[%
 aip,
 amsmath,amssymb,
 reprint,%
]{revtex4-1}

\usepackage{dcolumn}
\usepackage{bm}

\usepackage[utf8]{inputenc}
\usepackage[T1]{fontenc}
\usepackage{mathptmx}
\usepackage{etoolbox}
\usepackage{hyperref}       
\usepackage{url}            
\usepackage{booktabs}       
\usepackage{amsfonts}       
\usepackage{nicefrac}       
\usepackage{microtype}      
\usepackage{xcolor}         
\usepackage{graphicx}       
\graphicspath{ {./figures/} }
\usepackage{multirow}       
\usepackage[version=4]{mhchem}
\usepackage{float}
\makeatletter
\let\newfloat\newfloat@ltx
\makeatother
\usepackage{algorithm}
\usepackage{algorithmic}

\makeatletter
\def\@email#1#2{%
 \endgroup
 \patchcmd{\titleblock@produce}
  {\frontmatter@RRAPformat}
  {\frontmatter@RRAPformat{\produce@RRAP{*#1\href{mailto:#2}{#2}}}\frontmatter@RRAPformat}
  {}{}
}%
\makeatother
\begin{document}

\preprint{AIP/123-QED}

\title{Molecular Hypergraph Neural Networks}
\author{Junwu Chen}
\author{Philippe Schwaller}%
 \email{philippe.schwaller@epfl.ch}
\affiliation{ 
Laboratory of Artificial Chemical Intelligence (LIAC), Institute of Chemical Sciences and Engineering, Ecole Polytechnique F\'{e}d\'{e}rale de Lausanne (EPFL), Lausanne, Switzerland.
}
\affiliation{National Centre of Competence in Research (NCCR) Catalysis, Ecole Polytechnique F\'{e}d\'{e}rale de Lausanne (EPFL), Lausanne, Switzerland.
}

\date{\today}

\begin{abstract}
Graph neural networks (GNNs) have demonstrated promising performance across various chemistry-related tasks. However, conventional graphs only model the pairwise connectivity in molecules, failing to adequately represent higher-order connections like multi-center bonds and conjugated structures. To tackle this challenge, we introduce molecular hypergraphs and propose Molecular Hypergraph Neural Networks (MHNN) to predict the optoelectronic properties of organic semiconductors, where hyperedges represent conjugated structures. A general algorithm is designed for irregular high-order connections, which can efficiently operate on molecular hypergraphs with hyperedges of various orders. The results show that MHNN outperforms all baseline models on most tasks of OPV, OCELOTv1 and PCQM4Mv2 datasets. Notably, MHNN achieves this without any 3D geometric information, surpassing the baseline model that utilizes atom positions. Moreover, MHNN achieves better performance than pretrained GNNs under limited training data, underscoring its excellent data efficiency. This work provides a new strategy for more general molecular representations and property prediction tasks related to high-order connections.
\end{abstract}

\maketitle

\section{Introduction}

Graph presentation of molecular structures, also called molecular graphs, finds extensive application in computational chemistry and machine learning, where atoms are served as nodes and chemical bonds as edges. Graph neural networks (GNNs) are a class of deep learning models that can handle graph-structured data and are related to geometric deep learning \cite{duvenaud2015convolutional, gilmer2017neural, gasteiger2020directional, reiser_graph_2022, atz_geometric_2021}. Unlike traditional neural networks that operate on regular grids (e.g., images) or sequential data (e.g., text), GNNs can handle interconnected and non-Euclidean data, making them suitable for tasks involving graphs with complex topologies \cite{reiser_graph_2022}. This inherent advantage enables GNNs to directly learn the complex topological relationships of atoms and chemical bonds through molecular graphs \cite{fang_geometry-enhanced_2022}. In recent years, GNNs have demonstrated excellent molecular representation capabilities and achieved promising performance on many chemistry-related tasks, such as molecular property prediction \cite{yang2019analyzing, rampasek_recipe_2022, fang_geometry-enhanced_2022}, drug design \cite{yang_concepts_2019, li_graph_2022, sestak2023vn}, interatomic potentials \cite{smith2017ani, batatia2022mace,batzner_e3-equivariant_2022}, spectroscopic analysis \cite{mcgill2021predicting, singh2022graph, yang2021predicting}, reaction prediction and retrosynthesis \cite{coley2019graph, chen_generalized-template-based_2022, zhang2022chemistry}.

\begin{figure}
 \centering
 \includegraphics[width=0.48\textwidth]{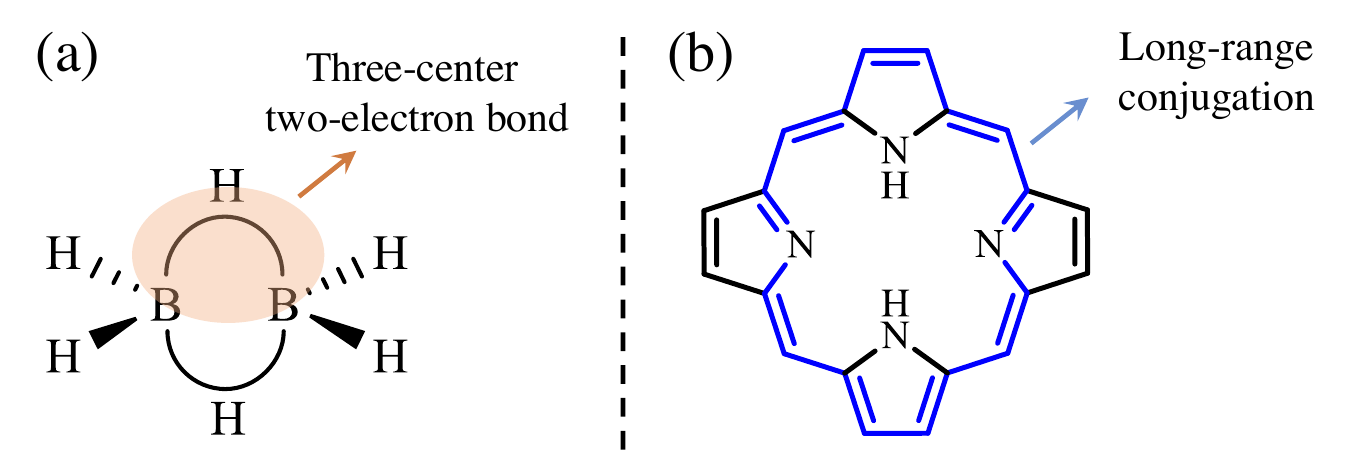}
 \caption{(a) Diborane structure and its 3-center-2-electron bond (B-H-B). (b) Porphyrin structure and its long-range conjugated bond.}
 \label{intro}
\end{figure}

However, ordinary graphs are limited to modeling pairwise connectivity within molecular structures, falling short in effectively representing higher-order connections \cite{konstantinova_molecular_1995, skvortsova_molecular_2021, sestak2023vn}. A substantial number of molecules have delocalized bonds, such as multi-center bonds \cite{szczepanik_uniform_2014} and conjugated bonds \cite{feixas_understanding_2011}. In contrast to classical chemical bonds localized between pairs of atoms, each delocalized bond involves three or more atoms \cite{merino_description_2005}. As illustrated in Figure~\ref{intro}a, two B atoms and one H atom share two electrons to form a 3-center-2-electron bond, which cannot be represented by a pairwise edge \cite{liao2012interpreting}. Similarly, conjugated organic molecules like porphyrin in Figure~\ref{intro}b, possess long-range dispersed $\pi$ electrons beyond the descriptive capability of conventional edges \cite{feixas_understanding_2011}. Therefore, the development of a more comprehensive graph representation for molecular structures becomes imperative to address this limitation inherent to conventional graphs.

A hypergraph is a generalization of the graph where a hyperedge can join any number of nodes \cite{bai_hypergraph_2021, antelmi_survey_2023}. Due to the innate ability to capture higher-order relationships, hypergraphs can powerfully model complex topological structures such as social networks \cite{aponte_hypergraph_2022}, chemical reactions \cite{schwaller2020predicting}, and compound–protein interactions \cite{saifuddin_hygnn_2023, murgas_hypergraph_2022, sestak2023vn}. Hypergraph Neural Networks (HGNs) belong to a category of neural networks designed to work with hypergraphs and extend the idea of GNNs to handle hyperedges \cite{antelmi_survey_2023, saifuddin_hygnn_2023}. Several studies \cite{kajino_molecular_2019,jo2021edge} have employed HGNs in the field of chemistry and depicted atoms as hyperedges and bonds between two atoms as nodes. While these approaches improve the validity of molecule generation and enhance edge representation learning \cite{kajino_molecular_2019,jo2021edge}, they presently do not leverage hyperedges to articulate high-order connections within molecules. For diverse molecular structures, especially organometallic complexes, and conjugated molecules, hyperedges from hypergraphs are competent to represent multi-atomic connections like delocalized bonds due to their inherent advantages \cite{konstantinova1995molecular, konstantinova2001application}.

\begin{figure*}
 \centering
 \includegraphics[width=\textwidth]{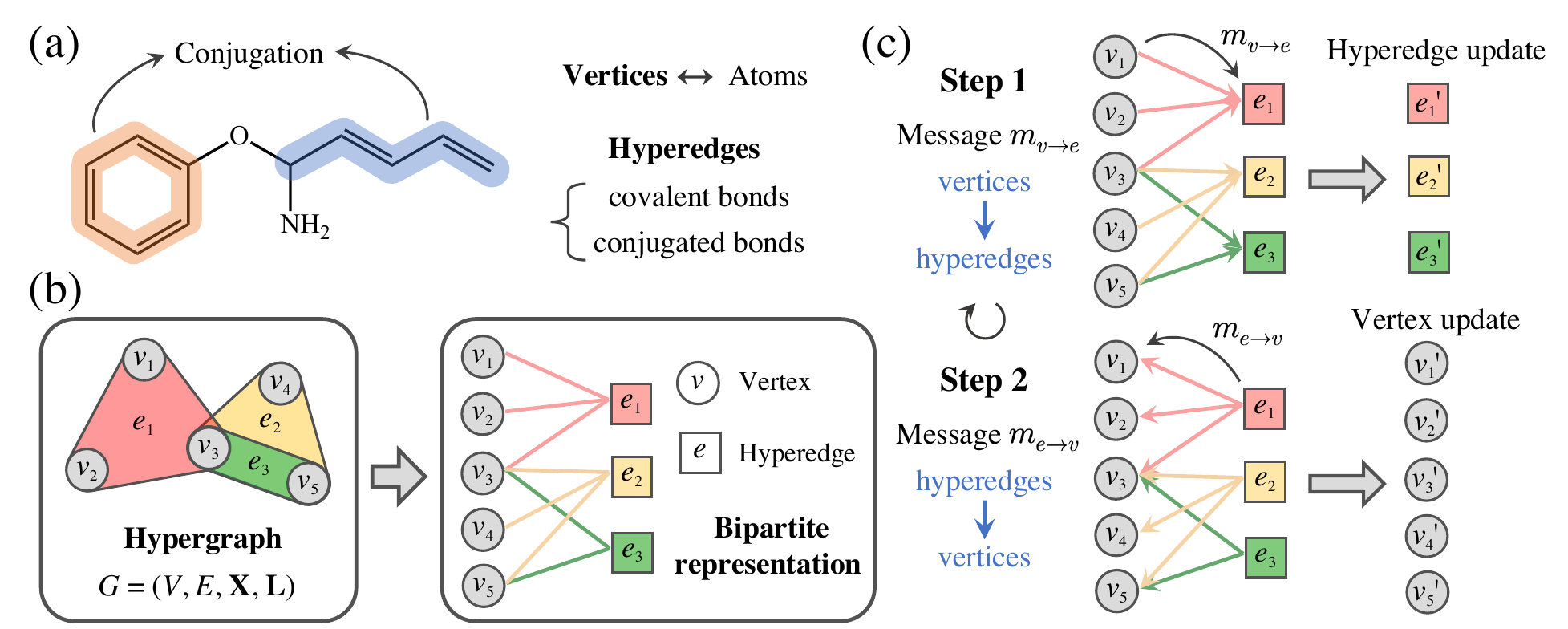}
 \caption{(a) The method of constructing molecular hypergraphs for conjugated molecules. (b) The conversion from a hypergraph to an equivalent bipartite graph. (c) The message passing method of our MHNN model.}
 \label{method}
\end{figure*}

Conjugated molecules, characterized by alternating single and multiple bonds along a molecular backbone, play a pivotal role in photoelectric applications such as organic light-emitting diodes (OLEDs) and organic solar cells (OSCs) \cite{dimitriev2022dynamics,bronstein2020role}. Their distinctive advantage stems from the delocalized $\pi$ electrons within conjugated structures, which can facilitate charge transport and optical absorption, establishing them as indispensable components of organic semiconductors \cite{bronstein2020role}. Although various machine learning models, especially GNNs, have been developed for predicting optoelectronic properties and accelerating the design of organic semiconductors \cite{st_john_message-passing_2019, bhat_electronic_2022, lu2020deep, nagasawa2018computer}, high-order conjugated connections have still not been properly modeled.

Herein, we introduce the concept of molecular hypergraphs and propose a Molecular Hypergraph Neural Network (MHNN) based on a simple but general message-passing method. MHNN was implemented to predict the optoelectronic properties of organic semiconductors where hyperedges represent conjugated structures. On three photovoltaic-related datasets, MHNN outperforms all baseline models in most tasks. Despite not using any 3D geometric information, MHNN exhibits better results than 3D-based models like SchNet \cite{schutt_schnet_2017} which require atom coordinates as input. Moreover, MHNN possesses high data efficiency even compared with pretrained models, which could be useful for data-scarce applications. This work provides a new model for property prediction of complex molecules containing higher-order connections.

\section{Methods}

\subsection{Molecular hypergraph}

A hypergraph $G=\left( V, E, \mathbf{H}, \mathbf{L}\right)$ is defined by a set of $n$ nodes $V$, a set of $m$ hyperedges $E$, node features $\mathbf{H} \in \mathbb{R}^{n \times d}$, and hyperedge features $\mathbf{L} \in \mathbb{R}^{m \times d^{\prime}}$. Each hyperedge $e = \left \{ v_1, \cdots, v_{|e|} \right \}$ is a subset of $V$ and its order $|e| \geq 2$. In a molecular hypergraph, it is natural to employ nodes to represent atoms and hyperedges to represent pairwise bonds, delocalized bonds, conjugated bonds and other higher-order associations. It is worth noting that the definition of hyperedges is important and should be related to the prediction target. For example, conjugated structures can significantly affect the light absorption and emission of molecules, so it is reasonable to describe conjugated bonds with hyperedges for the prediction of optoelectronic properties (e.g., bandgap).\cite{bronstein2020role} Moreover, hyperedges could be defined by pharmacophores \cite{yang2010pharmacophore} or toxicophores \cite{webel2020revealing} for the prediction of molecular activity or toxicity, respectively. In this work, we show an example of using molecular hypergraphs to describe conjugated molecules (Fig. \ref{method}a), where hyperedges are constructed by pairwise bonds and conjugated bonds. Like benzene (\ce{C6H6}) containing 12 atoms, six C-H $\sigma$ bonds, six C-C $\sigma$ bonds, and one large delocalized $\pi$ bond, its molecular hypergraph consists of twelve nodes, twelve 2-order hyperedges, and one 6-order hyperedge.

\subsection{Algorithm}

\begin{figure*}
    \centering
    \includegraphics[width=0.8\textwidth]{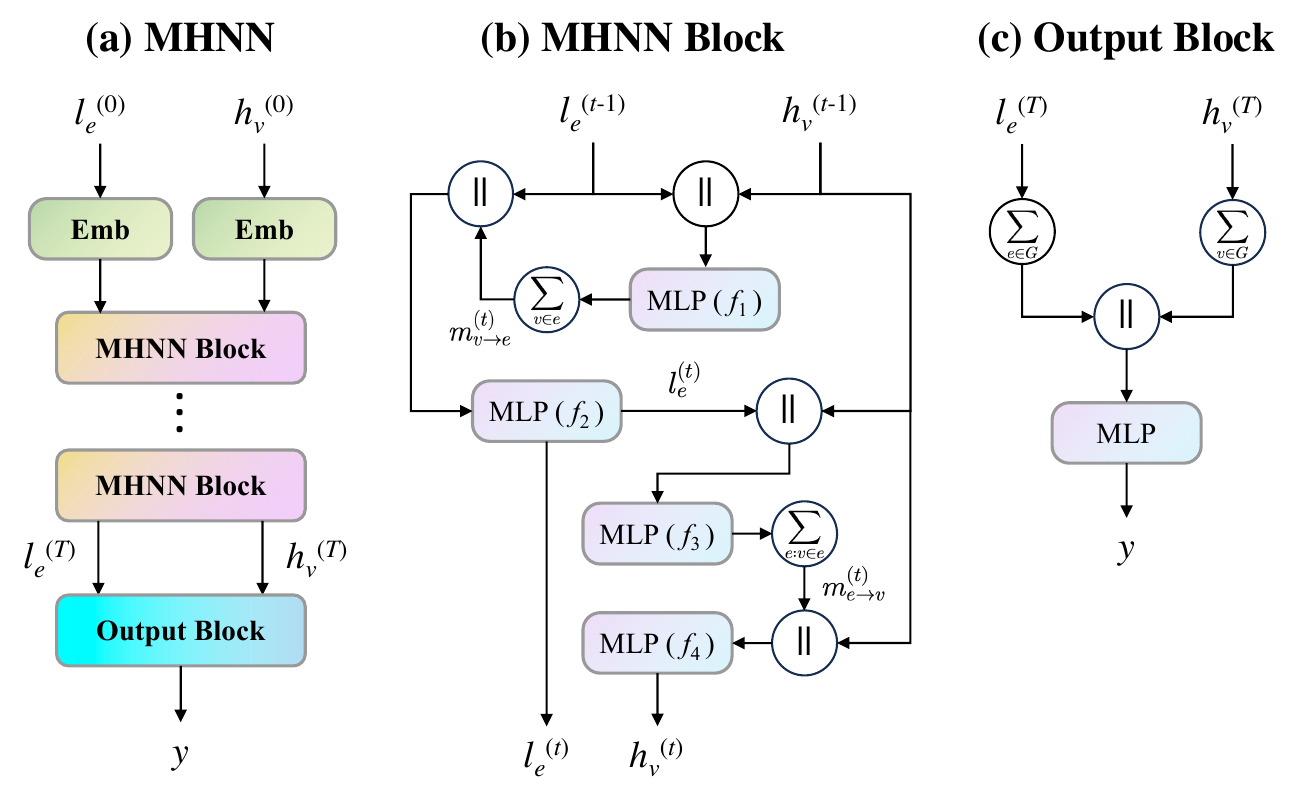}
    \caption{The MHNN architecture. $||$ denotes concatenation. The embeddings of nodes and hyperedges are updated in multiple MHNN blocks which can share parameters or not. The final embeddings of nodes and hyperedges are passed into an output block to generate predictions.}
    \label{mhnn}
\end{figure*}

The higher-order relations in complex molecules are often very diverse, that is, the orders of hyperedges in molecular hypergraphs often vary. For example, the number of atoms contained in a conjugated bond can be any integer greater than four. Therefore, model algorithms should not be limited to hyperedges of a specific order or within a specific order range. In addition, the model should also have good extrapolation ability for hyperedges of unseen orders. Inspired by recent works about hypergraph diffusion algorithms \cite{wang_equivariant_2022, wei_augmentations_2022}, we propose the Molecular Hypergraph Neural Networks (MHNN) based on bipartite representations of hypergraphs, which can efficiently operate on hypergraphs with hyperedges of various orders (Fig. \ref{method}bc).

The molecular hypergraph is initially transformed into an equivalent bipartite graph (Fig. \ref{method}b), wherein two distinct sets of vertices denote the nodes and hyperedges of the molecular hypergraph, respectively. The message passing of MHNN relies on the bipartite representations converted from molecular hypergraphs. Each message passing layer of MHNN is defined in terms of four differentiable functions $f_1$, $f_2$, $f_3$, and $f_4$. In the $t \  (1 \leq t \leq T)$ step message passing, the hidden states $l^{(t)}_{e}$ of each hyperedge are updated based on the messages $m^{(t)}_{v \to e}$ from the connected nodes ($v \in e$) according to:
\begin{align}
m^{(t)}_{v \to e} &= \sum_{v \in e} f_1\left( h^{(t-1)}_{v}, l^{(t-1)}_{e} \right) \\
l^{(t)}_{e} &= f_2\left( l^{(t-1)}_{e}, m^{(t)}_{v \to e} \right)
\end{align}
Then, the hidden states $h^{(t)}_{v}$ of each node are updated based on the messages $m^{(t)}_{e \to v}$ from involved hyperedges ($e:v \in e$) according to:
\begin{align}
m^{(t)}_{e \to v} &= \sum_{e:v \in e} f_3\left( l^{(t)}_{e}, h^{(t-1)}_{v} \right) \\
h^{(t)}_{v} &= f_4\left( h^{(t-1)}_{v}, m^{(t)}_{e \to v} \right)
\end{align}
where $h^{(0)}_{v}$ and $l^{(0)}_{e}$ are derived from initial atom features and bond features (Appendix \ref{ini-features}). After $T$ steps message passing, the hypergraph-level prediction is calculated in the readout part based on the final hidden states of nodes and hyperedges ($|e| > 2$), according to:
\begin{align}
\hat{y} = \operatorname{MLP}\left( \sum_{v \in G}h^{(T)}_v, \sum_{e \in G}l^{(T)}_e \right)
\end{align}
where $\operatorname{MLP}(\cdot )$ is a Multi-Layer Perceptron. The output $\hat{y}$ is the prediction target of MHNN, which can be a scalar or a vector. In this work, four $\operatorname{MLPs}$ are used to act as update functions ($f_1$, $f_2$, $f_3$, $f_4$). The schematic diagram of MHNN architecture is shown in Fig. \ref{mhnn} and Algorithm \ref{alg1}. 

\begin{algorithm}
\renewcommand{\algorithmicrequire}{\textbf{Input:}}
\renewcommand{\algorithmicensure}{\textbf{Output:}}
\caption{Algorithm of MHNN}
\label{alg1}
\begin{algorithmic}[1]
\REQUIRE molecular hypergraph $G=\left( V, E, \mathbf{H}, \mathbf{L}\right)$
\STATE Initialization: four MLPs ($f_1, f_2, f_3, f_4$) in each MHNN block, which can share parameters across $T$ layers or not. One MLP in the output block.
\FOR{$t=1,2,...,T$}
\STATE Send messages from $V$ to $E$ for all $e \in E$: \\ $m^{(t)}_{v \to e} = \sum_{v \in e} f_1\left( \left[ h^{(t-1)}_{v}, l^{(t-1)}_{e}\right] \right)$
\STATE Update hyperedge embeddings $l^{(t)}_{e} = f_2\left( \left[ l^{(t-1)}_{e}, m^{(t)}_{v \to e} \right] \right)$
\STATE Send messages from $E$ to $V$: $m^{(t)}_{e \to v} = \sum_{e:v \in e} f_3\left( \left[ l^{(t)}_{e}, h^{(t-1)}_{v} \right] \right)$
\STATE Update node embeddings $h^{(t)}_{v} = f_4\left( \left[ h^{(t-1)}_{v}, m^{(t)}_{e \to v} \right] \right)$
\ENDFOR
\STATE hypergraph embedding from nodes: $g_v = \sum_{v \in G}h^{(T)}_v$
\STATE hypergraph embedding from hyperedges: $g_e =\sum_{e \in G}l^{(T)}_e, \  |e|>2$
\STATE $\hat{y} = \operatorname{MLP}\left( \left[ g_v, g_e \right] \right)$
\ENSURE $\hat{y}$
\end{algorithmic}
\end{algorithm}

\begin{figure*}
\includegraphics[width=\textwidth]{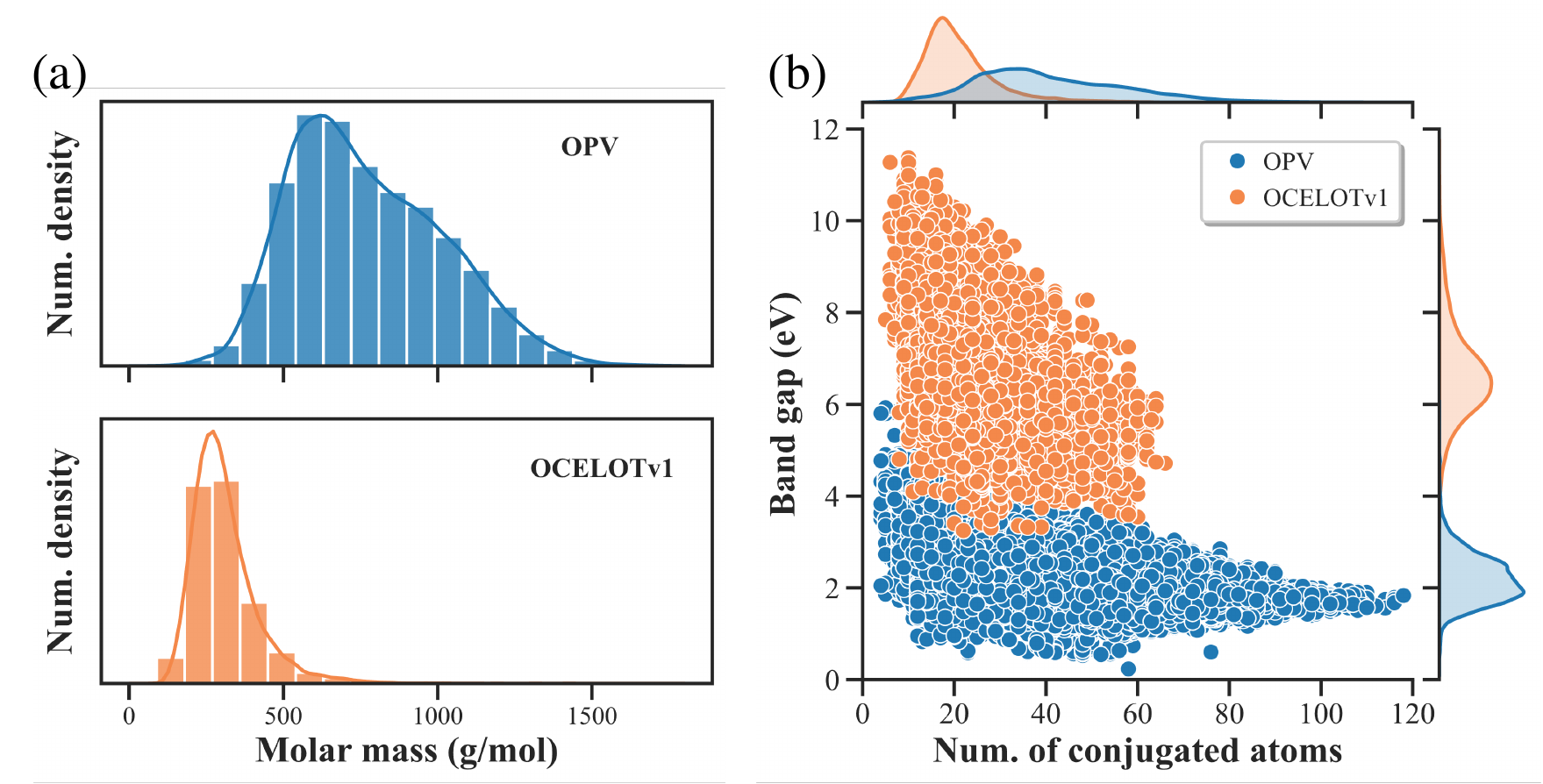}
\caption{\label{fig:datasets}(a) Distribution of molecular weights for OPV and OCELOTv1 datasets. (b) Distribution of band gap and atomic number of conjugated structures for OPV and OCELOTv1 datasets.}
\end{figure*}

\subsection{Input features}
For 2D GNN baselines, the atoms features and bond features designed by OGB \cite{hu_open_2020} are used for the initial features of models. For MHNN, initial atom features are from OGB \cite{hu_open_2020} and only bond types are used as the initial feature of all hyperedges. For 3D GNN baselines, only atomic numbers are used as the initial node feature. More details are listed in the Appendix \ref{ini-features}.

\subsection{Datasets}

\begin{table}[htp]
\caption{\label{tab:datasets}Overview of the datasets }
\begin{ruledtabular}
\begin{tabular}{lcccc}
Dataset & Graphs & Task type & Task number & Metric \\
\hline
OPV & 90,823 & regression & 8 & MAE  \\
OCELOTv1 & 25,251 & regression & 15 & MAE  \\
PCQM4Mv2 & 3,746,620 & regression & 1 & MAE  \\
\end{tabular}
\end{ruledtabular}
\end{table}

The OPV dataset \cite{st_john_message-passing_2019}, named organic photovoltaic dataset, contains 90,823 unique molecules (monomers and soluble small molecules) and their SMILES strings, 3D geometries, and optoelectronic properties from DFT calculations. OPV has four molecular tasks for monomers, the energy of highest occupied molecular orbital ($\varepsilon_{\rm HOMO}$), lowest unoccupied molecular orbital ($\varepsilon_{\rm LUMO}$), HOMO-LUMO gap ($\Delta \varepsilon$), and the spectral overlap $I_{overlap}$. In addition, OPV has four polymeric tasks, the polymer $\varepsilon_{\rm HOMO}$, polymer $\varepsilon_{\rm LUMO}$, polymer gap $\Delta \varepsilon$, and optical LUMO $O_{\rm LUMO}$. \cite{st_john_message-passing_2019}

The OCELOTv1 dataset \cite{bhat_electronic_2022} comprises about 25,000 organic $\pi$-conjugated molecules, along with their optoelectronic and reaction characteristics calculated by precise DFT or TD-DFT methods. The dataset encompasses 15 molecular properties: vertical (VIE) and adiabatic (AIE) ionization energy, vertical (VEA) and adiabatic (AEA) electron affinity, cation (CR) and anion (AR) relaxation energy, HOMO and LUMO energy, HOMO–LUMO energy gap (H–L), electron (ER) and hole (HR) reorganization energy, and lowest-lying singlet (S0S1) and triplet (S0T1) excitation energy. 

PCQM4Mv2 \cite{hu_ogb-lsc_2021} is based on the PubChemQC project \cite{nakata_pubchemqc_2017} and aims to predict the HOMO-LUMO energy gap of molecules from SMILES strings. PCQM4Mv2 is unprecedentedly large (> 3.8M graphs) compared to other labeled graph-related databases.

We follow the standard train/validation/test dataset splits from OPV and PCQM4Mv2, and use random split for the OCELOT dataset. The experimental results are derived from three separate runs using different random seeds, except for PCQM4Mv2, which is based on one single random seed run.

\begin{table*}
\caption{\label{table1}MAE results on OPV testing set. The unit of $I_{overlap}$ target is W/mol, and the unit of other targets is meV. * represents using DFT-optimized atom coordinates during model training. The results of MPNN and SchNet are from the reference \cite{st_john_message-passing_2019}.}
\begin{ruledtabular}
\begin{tabular}{l *8{c}}
    \multirow{2.3}{*}{Methods} & \multicolumn{4}{c}{Molecular} & \multicolumn{4}{c}{Polymer} \\
    \cmidrule(lr){2-5} \cmidrule(lr){6-9}
    & $\Delta \varepsilon$ & $\varepsilon_{\rm HOMO}$ & $\varepsilon_{\rm LUMO}$ & $I_{overlap}$ & $\Delta \varepsilon$ & $\varepsilon_{\rm HOMO}$ & $\varepsilon_{\rm LUMO}$ & $O_{\rm LUMO}$ \\
    \midrule
    GCN & 67.9 $\pm$ 1.2 & 38.2 $\pm$ 0.3 & 55.3 $\pm$ 1.5 & 265.8 $\pm$ 4.4 & 76.2 $\pm$ 1.4 & 54.2 $\pm$ 0.5 & 61.8 $\pm$ 0.6 & 61.6 $\pm$ 0.5      \\
    GIN & 48.5 $\pm$ 0.4 & 29.2 $\pm$ 0.2 & 38.6 $\pm$ 0.6 & 188.8 $\pm$ 2.8 & 66.8 $\pm$ 0.7 & 48.8 $\pm$ 0.4 & 54.9 $\pm$ 0.6 & 54.0 $\pm$ 0.3      \\
    GAT & 54.7 $\pm$ 1.2 & 33.5 $\pm$ 0.7 & 42.9 $\pm$ 1.6 & 204.2 $\pm$ 7.7 & 72.5 $\pm$ 1.7 & 51.6 $\pm$ 1.2 & 58.6 $\pm$ 0.8  &  58.0 $\pm$ 0.6  \\
    GATv2 & 57.7 $\pm$ 2.3 & 32.6 $\pm$ 1.5 & 44.0 $\pm$ 2.2 & 200.1 $\pm$ 1.3  &  73.1 $\pm$ 0.8 & 51.9 $\pm$ 0.3  & 57.8 $\pm$ 0.8 &  58.2 $\pm$ 0.8  \\
    MPNN & 36.9 $\pm$ 0.4 & 32.1 $\pm$ 0.8 & 27.9 $\pm$ 0.7 & 149.3 $\pm$ 2.3 & 57.1 $\pm$ 0.5 & 49.1 $\pm$ 0.8 & 47.8 $\pm$ 0.7 & 47.8 $\pm$ 0.5     \\
    SchNet* & 32.7 $\pm$ 0.5  & 27.0 $\pm$ 0.4 & 24.8 $\pm$ 0.4 & \textbf{96.6 $\pm$ 0.9} & 69.8 $\pm$ 0.6 & 56.9 $\pm$ 0.3 & 56.8 $\pm$ 0.5 & 57.2 $\pm$ 0.3    \\
    \midrule
    MHNN & \textbf{28.6 $\pm$ 0.2}  & \textbf{22.1 $\pm$ 0.1} & \textbf{21.2 $\pm$ 0.3} & 113.5 $\pm$ 0.7 & \textbf{56.6 $\pm$ 0.1} & \textbf{45.8 $\pm$ 0.7} & \textbf{45.1 $\pm$ 0.3} & \textbf{44.7 $\pm$ 0.1} \\
\end{tabular}
\end{ruledtabular}
\end{table*}

\section{Results and Discussion}

In this section, we initially assessed the predictive performance of MHNN on optoelectronic properties across three datasets. Among them, the OPV \cite{st_john_message-passing_2019} and OCELOTv1 \cite{st_john_message-passing_2019} datasets consist of conjugated molecules and their optoelectronic properties, while the PCQM4Mv2 dataset was employed to investigate the large-scale learning capability of MHNN. Subsequently, we explored the data efficiency of MHNN at different training data sizes.

\subsection{Analysis of datasets}
OPV and OCELOTv1 datasets, composed of conjugated molecules, are utilized to explore the learning ability of MHNN on conjugated structure and its prediction performance for optoelectronic properties. As shown in Fig. \ref{fig:datasets}a, the conjugated molecules in the OPV dataset have a broader molar mass distribution (80-1800 g/mol) compared to the OCELOTv1 dataset (90-1400 g/mol). The molecular weights in the OPV dataset are predominantly concentrated in the range of 500 to 1000, whereas the OCELOTv1 dataset shows a concentration in the range of 200 to 400. Therefore, the OPV dataset not only has more data points than the OCELOTv1 dataset, but also has more large conjugated molecules. As depicted in Fig. \ref{fig:datasets}b, molecules with larger conjugated structures are present in the OPV dataset compared to the OCELOTv1 dataset. The number of atoms in each conjugated structure of the OPV dataset spans a range from 4 to 120, with a concentration between 25 and 50. In contrast, the OCELOTv1 dataset exhibits a narrower range of atom numbers of conjugated structures (5-66), and is mainly concentrated between 15 and 30. Moreover, the conjugated molecules in the OPV dataset generally have lower band gaps ($\sim$ 1.9 eV) compared to the OCELOTv1 dataset ($\sim$ 6.2 eV). It can be concluded from Fig. \ref{fig:datasets}b that molecules with larger conjugated structures tend to have smaller band gaps, but this is not absolute. The distribution without obvious regularity also demonstrates the complex relationship between the photoelectric properties and conjugated structures. This underscores the significance of utilizing hyperedges to represent conjugated structures.

\subsection{Performance on OPV dataset}

For OPV dataset, we compared MHNN with multiple baselines: GCN \cite{kipf_semi-supervised_2017}, GIN \cite{xu_how_2019}, GAT \cite{velickovic_graph_2018}, GATv2 \cite{brody_how_2022}, MPNN \cite{gilmer_neural_2017} and SchNet \cite{schutt_schnet_2017}. Table~\ref{table1} shows the test performances of MHNN and competitive baselines on the OPV dataset, where the best results are marked in bold. Except for SchNet \cite{schutt_schnet_2017} which uses the 3D molecular geometries from DFT calculations, other models including MHNN, only use 2D topology information from SMILES strings. As for molecular properties, SchNet is obviously better than the 2D baselines, since 3D information is crucial for these properties \cite{st_john_message-passing_2019}. However, MHNN outperforms all baselines on three tasks ($\Delta \varepsilon$, $\varepsilon_{\rm HOMO}$, $\varepsilon_{\rm LUMO}$) without any 3D information, indicating that molecular hypergraphs with additional conjugation information are reliable representations of organic semiconductors. The SchNet model outperforms other models significantly in the prediction of the target $I_{overlap}$, indicating that the 3D molecular geometries can provide crucial and unique insights for predicting this target. For polymer property prediction tasks, SchNet \cite{schutt_schnet_2017} cannot exhibit better performance because only atom positions of monomers are available. It also suggests that polymer properties could be less dependent on the precise 3D structures of monomers \cite{st_john_message-passing_2019}. Overall, MHNN achieves the best results on 7 out of 8 tasks compared to baselines, which demonstrates the significance of molecular hypergraphs and the excellent performance of MHNN for property prediction of conjugated molecules.

\begin{table*}
\caption{\label{OCELOT}MAE results of baselines and MHNN on OCELOTv1 testing set. The unit of all targets is eV. The results of baselines are from the reference \cite{bhat_electronic_2022}.}
\begin{ruledtabular}
\begin{tabular}{lccccccc}
    Target & RR & SVM & KRR & FFN & MPNN & MPNN+MolDes & \textbf{MHNN} \\
    \hline
    HOMO & 0.345 $\pm$ 0.005 & 0.317 $\pm$ 0.003 & 0.337 $\pm$ 0.003 & 0.354 $\pm$ 0.012 & 0.796 $\pm$ 0.446 & 0.330 $\pm$ 0.028 & \textbf{0.306 $\pm$ 0.004 } \\
    LUMO & 0.340 $\pm$ 0.006 & 0.277 $\pm$ 0.005 & 0.306 $\pm$ 0.002 & 0.297 $\pm$ 0.004 & 0.291 $\pm$ 0.044 & 0.289 $\pm$ 0.028 & \textbf{0.258 $\pm$ 0.003} \\
    H-L  & 0.580 $\pm$ 0.005 & 0.604 $\pm$ 0.006 & 0.561 $\pm$ 0.004 & 0.578 $\pm$ 0.011 & 1.264 $\pm$ 0.696 & 0.548 $\pm$ 0.029 & \textbf{0.519 $\pm$ 0.011} \\
    VIE  & 0.231 $\pm$ 0.004 & 0.204 $\pm$ 0.002 & 0.241 $\pm$ 0.004 & 0.219 $\pm$ 0.001 & 0.202 $\pm$ 0.043 & 0.191 $\pm$ 0.024 & \textbf{0.178 $\pm$ 0.003} \\
    AIE  & 0.222 $\pm$ 0.002 & 0.193 $\pm$ 0.002 & 0.222 $\pm$ 0.004 & 0.207 $\pm$ 0.003 & 0.176 $\pm$ 0.015 & 0.173 $\pm$ 0.006 & \textbf{0.162 $\pm$ 0.004} \\
    CR1  & 0.058 $\pm$ 0.001 & 0.059 $\pm$ 0.001 & 0.057 $\pm$ 0.001 & 0.063 $\pm$ 0.001 & 0.054 $\pm$ 0.001 & 0.055 $\pm$ 0.002 & \textbf{0.053 $\pm$ 0.001} \\
    CR2  & 0.059 $\pm$ 0.001 & 0.061 $\pm$ 0.001 & 0.056 $\pm$ 0.001 & 0.059 $\pm$ 0.001 & 0.061 $\pm$ 0.001 & 0.053 $\pm$ 0.001 & \textbf{0.052 $\pm$ 0.000} \\
    HR   & 0.112 $\pm$ 0.001 & 0.114 $\pm$ 0.001 & 0.113 $\pm$ 0.001 & 0.110 $\pm$ 0.002 & 0.126 $\pm$ 0.022 & 0.133 $\pm$ 0.019 & \textbf{0.099 $\pm$ 0.001} \\
    VEA  & 0.218 $\pm$ 0.004 & 0.172 $\pm$ 0.002 & 0.231 $\pm$ 0.004 & 0.186 $\pm$ 0.002 & 0.193 $\pm$ 0.052 & 0.157 $\pm$ 0.018 & \textbf{0.138 $\pm$ 0.001} \\
    AEA  & 0.210 $\pm$ 0.001 & 0.182 $\pm$ 0.002 & 0.219 $\pm$ 0.002 & 0.176 $\pm$ 0.002 & 0.160 $\pm$ 0.027 & 0.154 $\pm$ 0.027 & \textbf{0.124 $\pm$ 0.002} \\
    AR1  & 0.057 $\pm$ 0.001 & 0.053 $\pm$ 0.001 & 0.057 $\pm$ 0.001 & 0.062 $\pm$ 0.002 & 0.057 $\pm$ 0.002 & 0.051 $\pm$ 0.001 & \textbf{0.050 $\pm$ 0.001} \\
    AR2  & 0.052 $\pm$ 0.001 & 0.051 $\pm$ 0.001 & 0.053 $\pm$ 0.000 & 0.051 $\pm$ 0.001 & 0.048 $\pm$ 0.002 & 0.052 $\pm$ 0.001 & \textbf{0.046 $\pm$ 0.001} \\
    ER   & 0.104 $\pm$ 0.020 & 0.099 $\pm$ 0.002 & 0.105 $\pm$ 0.002 & 0.101 $\pm$ 0.002 & 0.093 $\pm$ 0.002 & 0.098 $\pm$ 0.006 & \textbf{0.092 $\pm$ 0.001} \\
    S0S1 & 0.307 $\pm$ 0.006 & 0.275 $\pm$ 0.004 & 0.307 $\pm$ 0.002 & 0.282 $\pm$ 0.003 & 0.252 $\pm$ 0.017 & 0.249 $\pm$ 0.013 & \textbf{0.241 $\pm$ 0.003} \\
    S0T1 & 0.230 $\pm$ 0.003 & 0.183 $\pm$ 0.003 & 0.235 $\pm$ 0.004 & 0.194 $\pm$ 0.003 & 0.148 $\pm$ 0.012 & 0.150 $\pm$ 0.028 & \textbf{0.145 $\pm$ 0.002} \\
\end{tabular}
\end{ruledtabular}
\end{table*}

\subsection{Performance on OCELOTv1 dataset}

All models from the original paper \cite{bhat_electronic_2022} were selected as baseline models to compare the performance of MHNN on the OCELOTv1 dataset. Extended connectivity fingerprint (ECFP2) and 266 molecular descriptors were calculated from SMILES strings and used as the input for ridge regression (RR), support vector machine (SVM), kernel ridge regression (KRR) and feed-forward network (FFN) \cite{bhat_electronic_2022}. For the MPNN+MolDes model, the graph embeddings computed by MPNN are concatenated with the vectors of molecular descriptors, and employed for predicting molecular properties through a FFN \cite{bhat_electronic_2022}. More details about the baseline models can be found in Reference \cite{bhat_electronic_2022}. Table~\ref{OCELOT} shows the test performances of MHNN and baselines, where the best results are marked in bold. On the tasks such as AIE, AEA, S0S1 and S0T1, MPNN exhibits better performance than models (RR, SVM, KRR, FFN) using molecular descriptors. However, the models using molecular descriptors show superior performance than MPNN in the tasks like HOMO, H-L and HR. Moreover, with the assistance of extra molecular descriptors, MPNN+MolDes model demonstrates greater predictive performance across most tasks compared to other models. It indicates that both molecular graphs and molecular descriptors can provide important and specific information for the optoelectronic property prediction, respectively. Despite not using molecular descriptors, MHNN outperforms all baseline models in 15 tasks, demonstrating its excellent prediction performance. This illustrates that molecular hypergraphs are strong representations of conjugated molecules and MHNN can extract important information related to optoelectronic properties from conjugated structures.
 
\subsection{Performance on PCQM4Mv2 dataset}

\begin{table}[htp]
  \caption{Validate MAE results of MHNN and other message-passing GNN baselines on the PCQM4Mv2. The results of baselines are from the reference \cite{hu_ogb-lsc_2021, kim_pure_2022}. This dataset does not publish its test set. VN represents the use of virtual nodes to improve performance.}
  \label{ogb}
  \centering
  \begin{ruledtabular}
  \begin{tabular}{lcc}
    Model & Parameters & Validate MAE (eV) \\
    \hline
    GCN & 2.0 M & 0.1379 \\
    GIN & 3.8 M & 0.1195 \\
    GAT & 6.7 M & 0.1302 \\
    GCN-VN & 4.9 M & 0.1153 \\
    GAT-VN & 6.7 M & 0.1192 \\
    \hline
    MHNN & 2.1 M & \textbf{0.1125} \\
  \end{tabular}
  \end{ruledtabular}
\end{table}

To explore the learning ability on large-scale dataset, MHNN is compared with GNN baselines with a message passing mechanism on the PCQM4Mv2 dataset (Table \ref{ogb}). It should be pointed out that there are a large number of small molecules without conjugated structures in this dataset, even though the prediction target is band gap, one of the optoelectronic properties. As shown in Table \ref{ogb}, MHNN can obtain lower MAE results with fewer model parameters, which proves its high learning efficiency. This also shows that MHNN has reliable large-scale learning ability and could reduce the training cost on huge datasets.

\begin{figure*}
    \centering
    \includegraphics[width=\textwidth]{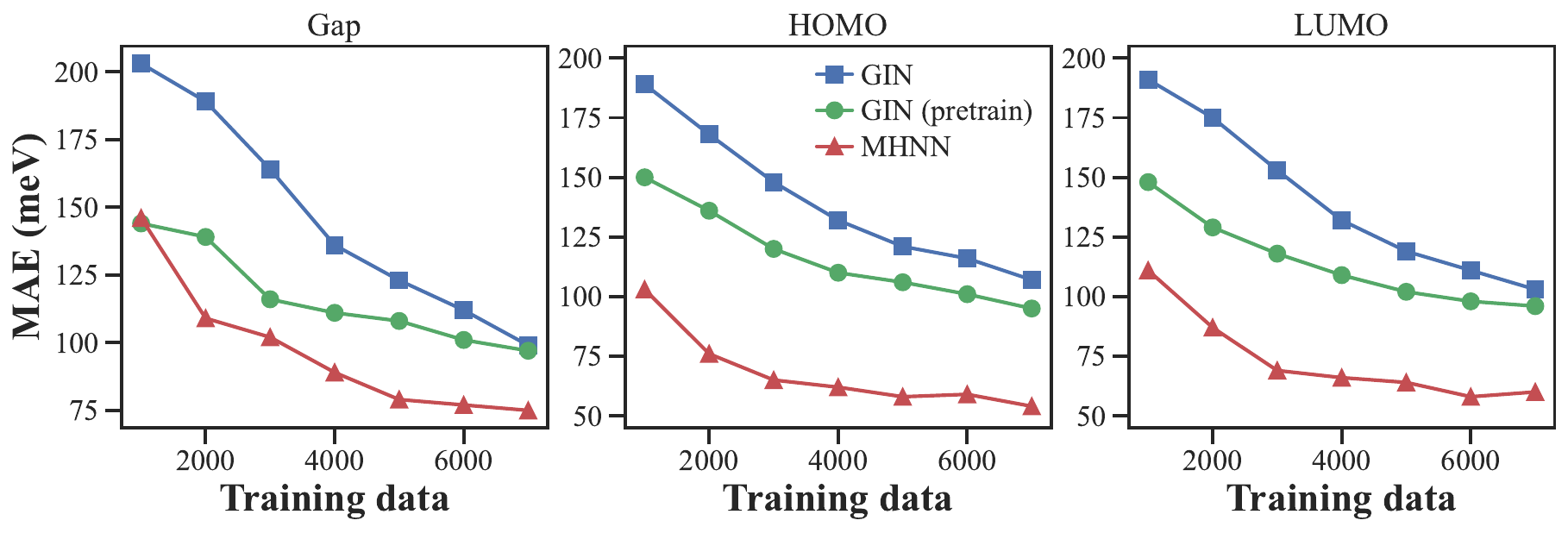}
    \caption{The test results of different models on the HOMO-LUMO gap, HOMO and LUMO tasks of OPV dataset under different amounts of training data. The green lines represent the results of pretrained GIN by self-supervised learning \cite{zhang_graph_2022}, while the blue and red lines show the results from GIN and MHNN without pretraining, respectively. Except for the MHNN model, all data are from the reference \cite{zhang_graph_2022}.}
    \label{data_efficiency}
\end{figure*}

\subsection{Data efficiency}

To explore the data efficiency of MHNN, we compare it to GIN with or without pretraining on the three most important tasks of OPV dataset under the same data partition. All 80,823 unlabelled molecules in the training set were used to pretrain the GIN model using self-supervised learning (SSL) strategy \cite{zhang_graph_2022}. Different amounts of data were randomly selected from the training set to directly train GIN and MHNN or finetune the pretrained GIN. As shown in Figure~\ref{data_efficiency}, MHNN exhibits better results on three tasks than GIN and pretrained GIN at the different training data sizes. For instance, using 1000 labeled training data, MHNN surpasses pretrained GIN by 31\% and 25\% on the $\varepsilon_{\rm HOMO}$ and $\varepsilon_{\rm LUMO}$ tasks, respectively. In addition, directly-trained GIN needs 4$\sim$6 times more training data to attain performance equivalent to MHNN. All the results show that MHNN is highly data-efficient and could be useful for applications without abundant labeled data.

\section{Conclusion}
The molecular hypergraph and corresponding MHNN were designed to overcome the limitations of traditional molecular graphs when it comes to representing high-order connections within complex molecules. The photoelectric property prediction task of organic semiconductors was selected to evaluate its prediction performance. The definition of molecular hyperedges is specified to focus on conjugated structures of molecules, which relies on human knowledge of relevant connections rather than learning directly from data. Across all three datasets (OPV, OCELOTv1, PCQM4Mv2), MHNN exhibits superior performance to the baselines on most tasks. Impressively, even in the absence of 3D geometric information, MHNN surpasses SchNet which relies on atom positions. Moreover, MHNN demonstrates higher data efficiency compared to pretrained models, making it valuable for applications where labeled data is scarce.

\begin{acknowledgments}
This publication was created as part of NCCR Catalysis (grant number 180544), a National Centre of Competence in Research funded by the Swiss National Science Foundation.
\end{acknowledgments}

\section*{AUTHOR DECLARATIONS}
\subsection*{Conflict of Interest}
The authors have no conflicts to disclose.

\section*{Code availability}
The Python code of MHNN and baseline models for optoelectronic property prediction on the  OPV, OCELOTv1, and PCQM4Mv2 dataset can be found on GitHub at \href{https://github.com/schwallergroup/mhnn}{https://github.com/schwallergroup/mhnn}.

\appendix
\renewcommand\thefigure{\Alph{section}.\arabic{figure}}
\renewcommand\thetable{\Alph{section}.\arabic{table}}
\setcounter{figure}{0}
\setcounter{table}{0}

\section{Implementation Details}
Our implementation is based on PyTorch and PyG \cite{paszke_pytorch_2019,fey_fast_2019}. The code of 2D GNN baselines is from OGB \cite{hu_open_2020}. The experiments were conducted in a collaborative computing cluster setting, featuring diverse CPU and GPU architectures. This included a combination of NVidia V100 (32GB) and RTX3090 (24GB) GPUs. For a fair comparison, the same training recipe was used for all the models on the same dataset. For baseline models, the hyperparameters were adopted from references \cite{hu_ogb-lsc_2021,st_john_message-passing_2019}.

\section{Input features} \label{ini-features}
The Tables \ref{atomf}, \ref{bondf}, and \ref{hyperf} describe the input features for atoms, pair-wise edges, and hyperedges. 
\begin{table}[htp]
  \caption{Atom (node) features for MHNN and 2D GNN baselines.}
  \centering
  \label{atomf}
  \begin{ruledtabular}
  \begin{tabular}{l|r}
    Feature & Description \\
    \hline
    Atom type & type of atom (ex. C, N, O), by atomic number  \\
    Chirality & unspecified, tetrahedral CW/CCW, or other \\
    Degree & number of bonds the atom is involved in \\
    Formal charge & integer electronic charge assigned to atom \\
    Hydrogens & number of bonded hydrogen atoms \\
    Radical electrons & the number of unpaired electrons \\
    Hybridization & sp, sp2, sp3, sp3d, or sp3d2 \\
    Aromaticity & whether this atom is part of an aromatic system \\
    Is in ring & whether the atom is in a ring \\
  \end{tabular}
  \end{ruledtabular}
\end{table}

\begin{table}[htp]
  \caption{Bond (edge) features for 2D GNN baselines.}
  \centering
  \label{bondf}
  \begin{ruledtabular}
  \begin{tabular}{l|r}
    Feature & Description \\
    \hline
    Bond type & single, double, triple, or aromatic  \\
    Bond stereo & none, any, E/Z or cis/trans  \\
    Is conjugated & whether the bond is conjugated  \\
  \end{tabular}
  \end{ruledtabular}
\end{table}

\begin{table}[htp]
  \caption{Using bond type as the hyperedge feature of MHNN.}
  \centering
  \label{hyperf}
  \begin{ruledtabular}
  \begin{tabular}{c|r}
    Edge order & Feature \\
    \hline
    $= 2$ & bond type: single, double, triple, or aromatic  \\
    $> 2$ & conjugated bonds  \\
  \end{tabular}
  \end{ruledtabular}
\end{table}

\clearpage
\section*{REFERENCES}
\bibliography{references}

\end{document}